\documentclass[preprint2]{aastex}

\shorttitle{Intermediate-Mass Black Holes and LISA}
\shortauthors{Clifford M. Will}

\begin{document}


\title{On the Rate of Detectability of Intermediate-Mass Black-Hole
Binaries using LISA}


\author{Clifford M. Will}
\affil{
Department of Physics, McDonnell Center for the Space Sciences, \\
Washington University, St. Louis, MO 63130 \footnote{Permanent
address}\\
Groupe Gravitation Relativiste et Cosmologie (GR$\varepsilon$CO) \\
Institut d'Astrophysique de Paris, 75014 France }


\begin{abstract}
We estimate the rate at which the proposed space gravitational-wave
interferometer LISA could detect intermediate-mass black-hole
binaries, that is, binaries containing a black hole in the mass range
10 -- 100 $M_\odot$ orbiting a black hole in the mass range
100 -- 1000 $M_\odot$.  For one-year integrations leading up to the
innermost stable orbit, and a signal-to-noise ratio of 10, we estimate
a detection rate of only 1 per million years for $10 M_\odot/100
M_\odot$ binaries.  The estimate uses the method
of parameter estimation via matched filtering, incorporates a
noise curve for LISA established by the LISA Science Team
that is available online, and employs an IMBH formation rate model used
by \citet{miller02}.  We find that the detectable distance
is relatively insensitive to LISA arm lengths or acceleration noise, 
but is roughly
inversely proportional to LISA position errors, and 
varies
roughly as $T^{1/2}$, where $T$ is the integration time in years.  We also
show that, while all IMBH systems in this mass range may be detected in the
Virgo cluster up to 40 years before merger, none can be detected there
earlier than 400 years before merger.  
An extended LISA mission that enabled 10-year integrations could
detect IMBH systems at the Virgo cluster 1000 years before merger, and
systems in galactic globular clusters a million years before merger.
We
compare and contrast these estimates with earlier estimates by 
\citet{miller02}.
\end{abstract}




\section{Introduction and summary}

In recent years, intermediate-mass black holes (IMBH), holes with
masses between hundreds and thousands of solar masses, have attracted
considerable interest, both as relics of the evolution and
structure of globular clusters, and as possible sources of
gravitational radiation for both ground- and space-based laser
interferometers \citep{millercolbert}.  \citet{miller02} has estimated the
rate of detectable IMBH binaries by the proposed space antenna LISA to
be given by
\begin{eqnarray} 
R &\approx& 7 \times 10^{-3}
\left ( \frac{h}{0.7} \right )^3
\left ( \frac{f_{\rm tot}}{0.1} \right )
\left ( \frac{\mu}{10 M_\odot} \right )^{1/2} 
\nonumber \\
&& \times 
\left ( \frac{M_{\rm max}}{100 M_\odot} \right )^{3/2}
\left ( \ln \frac{M_{\rm max}}{M_{\rm min}} \right )^{-1}
{\rm yr}^{-1}
\,,
\label{rate0}
\end{eqnarray}
where $h$ is the Hubble parameter in units of $100 \,{\rm km \,s}^{-1}
\, {\rm Mpc}^{-1}$, $f_{\rm tot}$ is the total fraction of globular
clusters that have IMBH, $\mu$ and $M$ are the reduced mass and total
mass of the binary respectively, and 
$M_{\rm min}$ and $M_{\rm max}$ denote the range over which such IMBH
may exist in globular clusters.

A key ingredient in calculating this rate is the distance to which
LISA could detect binary IMBH inspirals for a given signal-to-noise
ratio (SNR).  \citet{miller02} used estimates of signal-to-noise
ratios for binary inspiral derived by 
\citet{flanhughes} from semi-analytic LISA noise curves.
We have recalculated this distance using standard
methods based on matched-filtering for binary inspiral, and using
an up-to-date LISA noise
curve available online.\footnote{The 
Sensitivity Curve Generator was originally written by
Shane Larson and may be found online at
http://www.srl.caltech.edu/\~\,shane/sensitivity/MakeCurve\\.html}
We find significantly smaller distances and rates reached for a given SNR and
with a significantly different dependence on mass than those
obtained by \citet{miller02}.  Figure 1 shows distances reached for SNR=10
and one year of integration leading up to merger, 
as a function of total mass $M$ and reduced mass
parameter $\eta=m_1m_2/M^2$ ($0 < \eta \le 1/4$).   
For $M=100 M_\odot$ and reduced mass $\mu=\eta M = 10 M_\odot$, 
we find a distance 
$D_L \approx 18 $ Mpc, 11 times smaller than Miller's equation
(19) (we assume that all masses are suitably redshifted).  
The rate of detection we find is given by  
\begin{eqnarray}
R &\approx&  1.0 \times 10^{-6} \left ( \frac{h}{0.7} \right )^3 
\left ( \frac{f_{\rm tot}}{0.1} \right ) 
\left ( \frac{\mu}{10 M_\odot} \right )^{19/8}
\nonumber \\
&& 
\times \left ( \frac{M_{\rm max}}{100 M_\odot} \right )^{13/4}
\left ( \ln \frac{M_{\rm max}}{M_{\rm min}} \right )^{-1}
{\rm yr}^{-1}
\,.
\label{rate2}
\end{eqnarray}

\begin{figure}[t]
\plotone{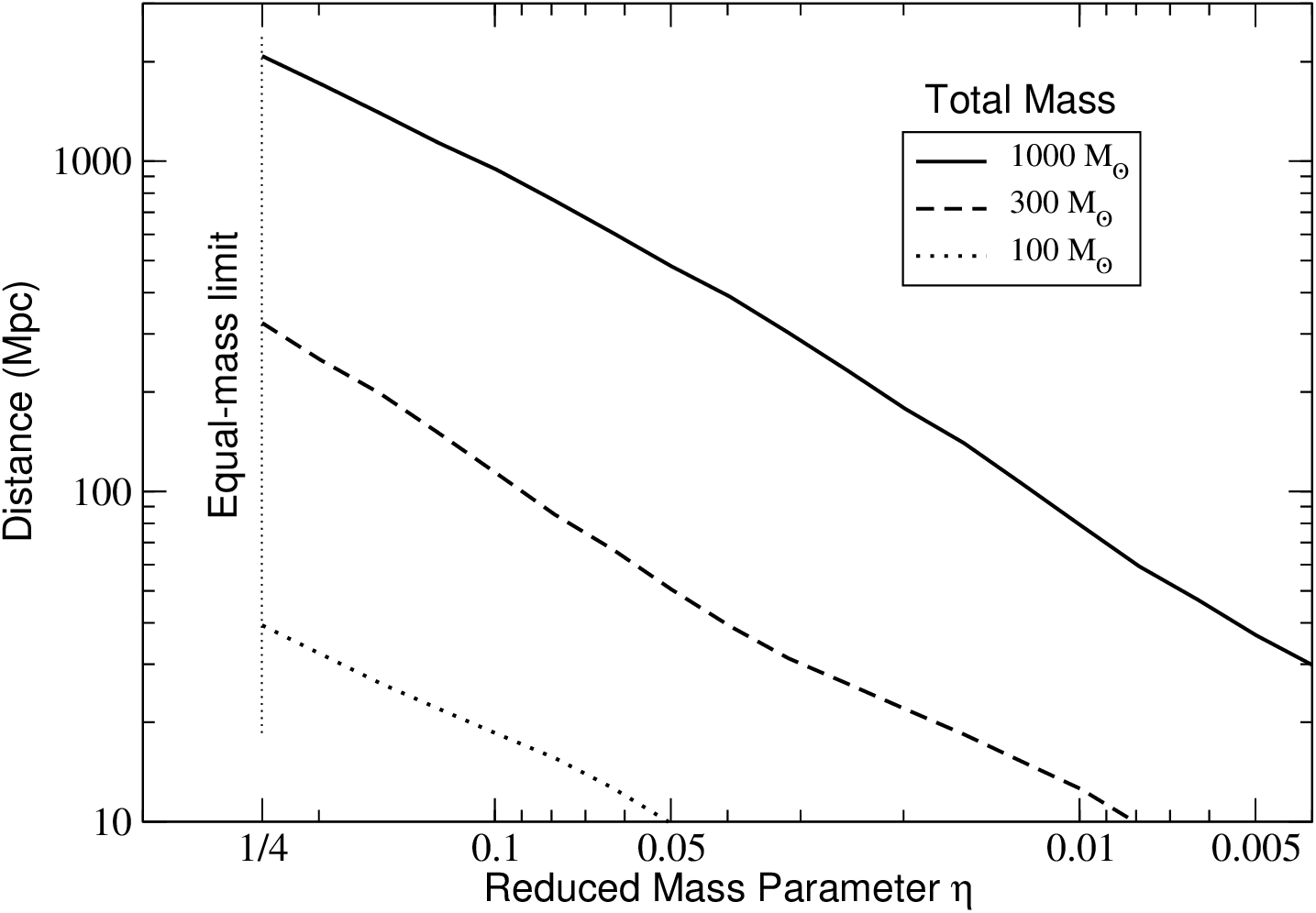}
\caption{Distances in Mpc 
reached by LISA for IMBH binary inspiral, for SNR=10 and
one year of integration prior to merger.}
\end{figure}

We also analyse LISA's reach for IMBH systems at epochs earlier than the
year leading up to merger.  Figure 2 shows distances as a function of years
before merger for various IMBH systems, for SNR=10 and one year of
integration.  While all systems in this mass range are detectable in the
Virgo cluster (at 19 Mpc) within 40 years of merger, {\em none} is
detectable there earlier than 400 years before merger, and at 1000 years
before merger, only systems closer than 7 Mpc are detectable.
On the other hand, we find that, for systems in this mass range, the
distance reached varies roughly as the square-root of the integration
time, so that an extended LISA mission that enabled 10-year
integrations, could reach roughly three times farther, 

\begin{figure}[t]
\plotone{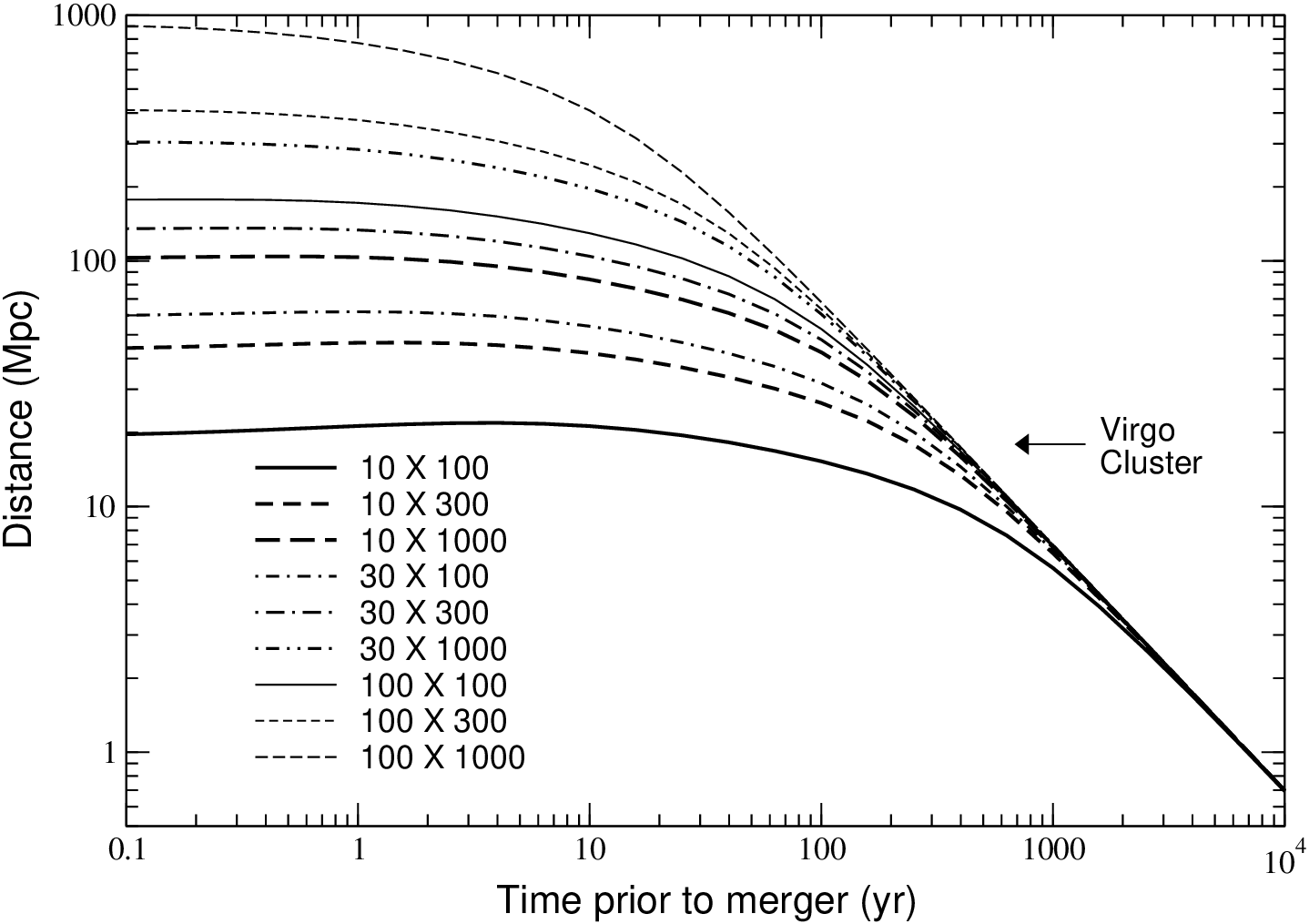}
\caption{Distances in Mpc
reached by LISA for IMBH binary inspiral, for SNR=10 and
one year of integration vs. years before merger, for various
black-hole pairs (in $M_\odot$).}
\end{figure}

The primary reason for this downward revision in the reach of LISA for IMBH
is our use of a more current sensivity curve for LISA, whose noise level is
substantially higher than the older Flanagan-Hughes curve used by Miller.
The rest of this paper gives the details to support this conclusion.
In \S 2 we determine the distance reached by LISA using standard SNR
calculations, and in \S 3 we use the method of \citet{miller02} to estimate
the detection rate of systems in the final year of inspiral.  In \S 4 we
discuss the results.

\section{Calculation of Distance Reached by LISA}

The signal-to-noise ratio $\rho$ for a gravitational-wave
signal whose Fourier transform is
$\tilde{h}(f)$ in a detector of noise spectral density $S_n(f)$ is
given by 
\begin{equation}
\rho[h]^2 \equiv 4 \int_{f_i}^{f_f} \frac{|\tilde{h}(f)|^2}{S_n(f)} df
\,,
\label{rho}
\end{equation}
where $f_i$ and $f_f$ are the initial and final frequencies between
which the signal is integrated (see 
\citet{3min,CutlerFlanagan,Finn,FinnChernoff,poissonwill} for details
of the method of matched filtering).  In the ``restricted post-Newtonian''
approximation, where the wave amplitude is given by the quadrupole
approximation and the wave phase is given by a high-order post-Newtonian 
expression, $\tilde{h}(f)$ is given by
\begin{equation}
\tilde{h}(f) = \left\{  \begin{array}{ll}
{\cal A} f^{-7/6} e^{i\Psi(f)} \,,  & 0<f<f_{\rm max} \,,\\
0  \,, & f>f_{\rm max} \,,
\end{array}  \right.
\label{fourier}
\end{equation}
where $f_{\rm max}$ is the largest frequency for which the wave can be
described by the restricted post-Newtonian approximation,
often taken to correspond to waves
emitted at the innermost stable orbit (ISCO) before the bodies
plunge toward each other and merge.  A useful approximation for this
frequency is (we use units in which $G=c=1$)
\begin{equation}
f_{\rm max} = (6^{3/2} \pi M)^{-1},
\label{iscofreq}
\end{equation}
where $M$ is the total mass of the system.
After averaging over all angles, the amplitude $\cal A$ is given by
\begin{equation}
{\cal A} = \frac{1}{\sqrt{30} \pi^{2/3}} \frac {{\cal M}^{5/6}}{D_L}
\,,
\label{amplitude}
\end{equation}
where 
$D_L$ is the luminosity
distance of the source, and ${\cal M} = \eta^{3/5}M$ is the ``chirp''
mass.  
In general relativity, the phasing function $\Psi(f)$ can be given to
high post-Newtonian order, but is not relevant for our purposes.
Combining equations (\ref{rho}), (\ref{fourier}) and (\ref{amplitude}), we
obtain the expression for $D_L$
\begin{equation}
D_L = \sqrt{\frac{2}{15}} \frac{{\cal M}^{5/6}}{\rho}
\frac{1}{\pi^{2/3}} B(7)^{1/2} \,,
\label{lumdistance}
\end{equation}
where
\begin{equation}
B(7) = \int_{f_i}^{f_f} \frac{f^{-7/3}}{S_n(f)} df \,.
\label{b7}
\end{equation}
For quasi-circular binary inspiral, the initial and final frequencies
can be related to the integration time $T$  using the expression
obtained from gravitational radiation damping of the orbit at
quadrupole order (which is sufficiently accurate for our purpose),
\begin{equation}
u_i = u_f \left ( 1 + {256 \over 5} {T \over {\cal M}} u_f^{8/3}
        \right )^{-3/8} \,,
\label{timeinterval}
\end{equation}
where $u=\pi {\cal M} f$.

\begin{figure}[t]
\plotone{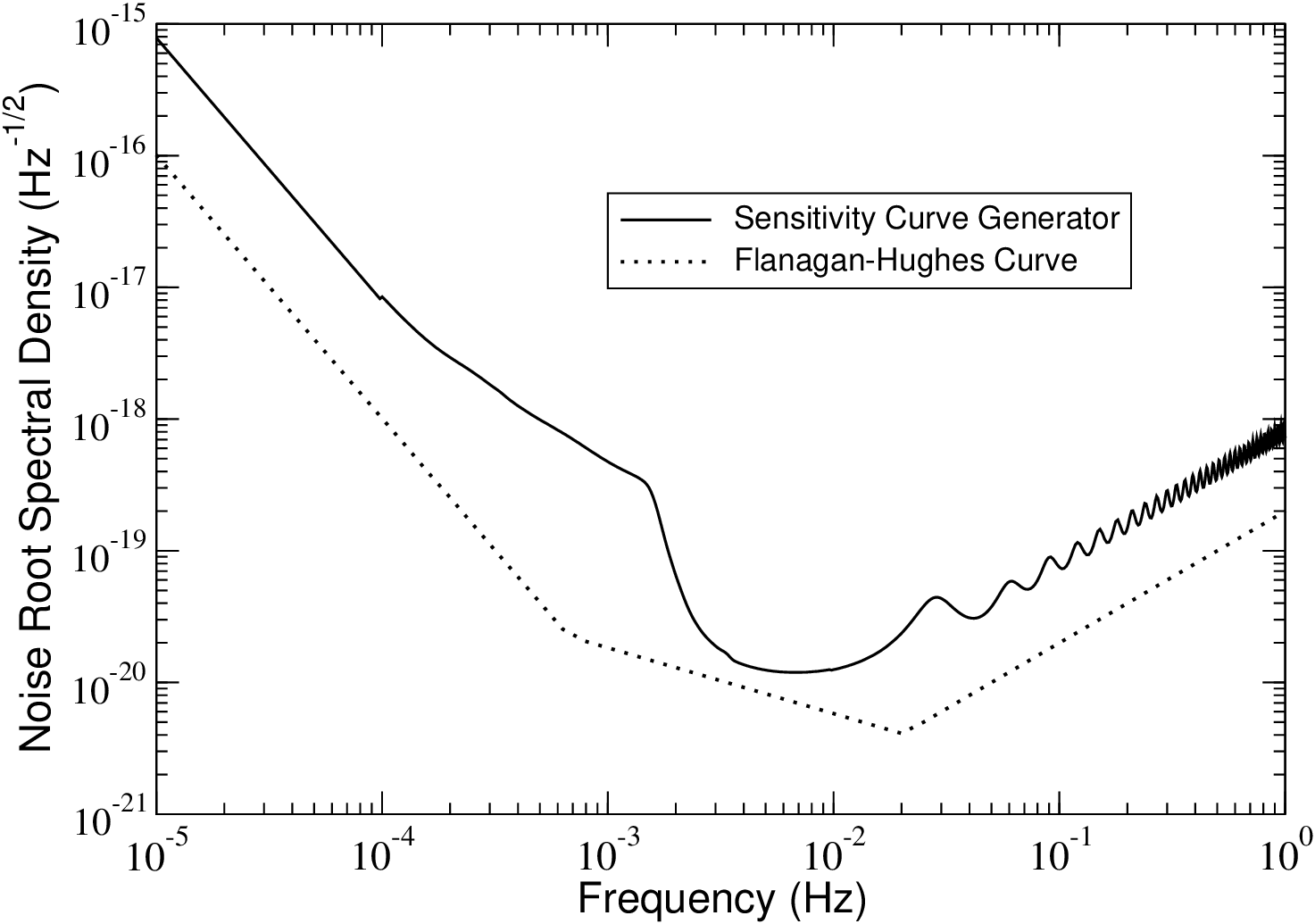}
\caption{Root noise spectral density for LISA}
\end{figure}

In this paper, we shall adopt sensitivity curves for LISA developed
independently by \citet{larson} and 
\citet{armstrong} (see figure 3).  The
two methods are in substantial agreement, and the former has been
summarized
in the Sensitivity Curve Generator (SCG), available online$^2$.
The sensitivity curves incorporate sources of
instrumental noise
in LISA, such as laser shot noise, acceleration noise, thermal noise,
etc.,
coupled with a LISA transfer function which takes into account the
effect of
the finite time of propagation of the laser beams during the passage
of the
gravitational waves.  We assume the case where all three arms of LISA
are of
equal length, and we assume
that averages over angles and polarizations have been done.
The SCG permits various choices to be made for LISA instrumental
parameters,
and has  an option to include
an estimate for confusion noise resulting from a background
of galactic white-dwarf binaries
\citep{benderhils}; this background is included in the analysis, but in
fact plays no significant role for the late stage
of IMBH inspiral because the signals are
predominantly at high frequency, well above the white-dwarf band.

We consider IMBH systems containing one black hole in the mass range 10 --
100 $M_\odot$ and a companion  black hole 
in the mass range 100 -- 1000 $M_\odot$.
Figure 1 shows the results for the distance as a function of reduced
mass parameter for various total masses, assuming
a SNR of 10 with a one-year integration time ending at the ISCO (or at
the termination of the LISA sensitivity curve, whichever comes first).  
The distances, in Mpc, can be
fit by the approximate formula
\begin{eqnarray}
D_L &=&  190 \left ( \frac{10}{\rho} \right )
\left ( \frac{{\cal M}}{100 M_\odot}  \right )^{15/8} 
\nonumber \\
&& \times 
\left ( \frac{T}{1 \, {\rm yr}}  \right )^{1/2} 
\, {\rm Mpc} 
\nonumber \\
&=&  40 \left ( \frac{10}{\rho} \right ) \left ( 4\eta
\right )^{9/8}
\left ( \frac{M}{100 M_\odot}  \right )^{15/8}
\nonumber \\
&& \times 
\left ( \frac{T}{1 \, {\rm yr}}  \right )^{1/2} 
\, {\rm Mpc} \,.
\label{lumdistance2}
\end{eqnarray}
This scaling can be understood as follows:  the waves from IMBH in the final
year of inspiral are
mainly at the high frequency end of the LISA sensitive band, where
$S_n(f) \propto f^2$.  Also, for these masses and year-long
integrations, the second term in equation (\ref{timeinterval}) dominates,
so that $f_i = (1/8\pi)(5/T)^{3/8}{\cal M}^{-5/8} \ll f_f$.  Consequently,
the 
integral $B(7) \sim {f_i}^{-10/3} \sim {\cal M}^{25/12} T^{5/4}$.
Substituting this into equation (\ref{lumdistance}), we obtain the scaling
in $\cal M$ shown in (\ref{lumdistance2}).  A  $T^{1/2}$
integration-time scaling actually fits the curves somewhat better than a
$T^{5/8}$ scaling inferred from the analytic estimate.
Equation (\ref{lumdistance2}) underestimates the distances at the low
mass end, e.g. for $\eta \sim 1/10$ and $M \sim 100 \, M_\odot$, by
about 20 percent.

It is straightforward to show that the distance reached
is independent of LISA acceleration noise, 
but is inversely proportional to the LISA position noise
error budget.  Halving the position noise doubles the distance
reached.  The distance is weakly dependent on LISA arm length, varying
by between 10 and 40 percent for factor-of-two changes in arm length;
varying arm length does not alter the overall level of the high-frequency end
of the noise curve, but instead moves the 
location of the peaks in the oscillations of the transfer function and
raises the floor of the noise curve near $10^{-3} \, {\rm Hz}$.

We obtain similar results using a sensitivity
curve derived by \citet{finnthorne} based on work of
the LISA Mission Definition team
(see reference 44 of \citet{finnthorne}).  This curve closely matches
the curve from the SCG except for the oscillations in the noise root
spectral density at
high frequency resulting from the LISA transfer function, an effect of 
the finite arm length.  The oscillations have the effect of reducing
somewhat the distances inferred from the SCG 
for a given SNR compared to those from the
Finn-Thorne curve, for sources at high frequency.

We also calculate the distances that can be reached in one year's
integration for IMBH systems much earlier than the ISCO stage (see Figure
2).  All systems
detected within about 40 years of merger, can be detected at the Virgo
cluster (19 Mpc) with SNR=10 and one year of integration.
However at earlier epochs the reach decreases dramatically, so that by 400
years {\em no} systems can be seen at Virgo, and by 
1000 years before merger, the distance reached for SNR=10 is in the range 5.6
to 7 Mpc for all IMBH mass ranges considered.  This is
well short of the Virgo cluster.  
For systems detected $10^6$ years before merger, the distance reached
is 7 Kpc, independent of mass.  This
behavior can be understood analytically.  For
observation epochs long before the ISCO, one can show that the inital and
final frequencies observed are related by $f_i \approx f_f
(1+T/T^\prime)^{-3/8}$, where $T$ is the integration time, and $T^\prime$ is
the epoch before merger, with $T \ll T^\prime$.  Then, in a region where the
noise spectral density is dominated by the 
behavior $S_n(f) \approx \alpha f^n$, one
can show, using equations (\ref{lumdistance}) and (\ref{b7}) that 
\begin{equation}
D_L \approx \frac{2}{5} \frac{(8\pi)^{n/2}}{\rho}
\left ( \frac{T}{\alpha} \right )^{1/2} 
\left ( \frac{T^\prime}{5} \right )^{\frac{3n-4}{16}} 
{\cal M}^{\frac{5(4+n)}{16}}
\,.
\end{equation}
For epochs earlier than
1000 years before merger, the systems are in the low-frequency
regime where $S_n(f) \propto f^{-4}$ (apart from white-dwarf confusion
noise), hence $D_L \approx (1/32\pi^2)(T /\alpha)^{1/2} (T^\prime \rho)^{-1}$, 
independent of chirp mass.  With $\alpha = 6.09 \times 10^{-51} \, {\rm
Hz}^3$ giving a good fit to the low-frequency LISA instrumental noise, we
obtain 
\begin{equation}
D_L \approx 7 \, T^{1/2} (1000/T^\prime)(10/\rho) \, {\rm Mpc} \,,
\end{equation}
which 
perfectly matches the large-epoch behavior in Figure 2.
The $T^{1/2}$ behavior is
expected for what are observations of an essentially CW source.  These
results also are far more pessimistic than the estimates made by
\citet{millercolbert}, which suggested
detecting IMBH binaries in the Virgo
cluster 1000 years before merger, and in the local system of globular
clusters $10^6$ years before merger.  On the other hand, because of
the $T^{1/2}$ behavior, such binaries {\em could} be reached in an
extended LISA mission that enabled 10-year integration times.

\section{Rate of detectable mergers}

To calculate the rate of inspirals 
detectable by LISA in the last year before merger we follow the
method outlined by \citet{miller02}.  The rate is given by
\begin{equation}
R = \frac{4\pi}{3} \int D(M)^3 \nu (M) n_{\rm gc} f(M) dM \,,
\label{rate1}
\end{equation}
where $D(M)$ is the distance reached for a given SNR  and integration
time, as a function of
total mass (holding reduced mass $\mu$ fixed); $\nu (M) = 10^{-10}
\mu^{-1} M \, {\rm yr}^{-1}$ is the rate at which smaller black holes
merge with black holes of mass $M$ in a given cluster; $n_{\rm gc} = 8h^3
\, {\rm Mpc}^{-3}$ is the number density of globular clusters in the
local universe; and $f(M) = [f_{\rm tot}/\ln (M_{\rm max}/M_{\rm
min})] M^{-1}$ is the fraction of globular clusters harboring black
holes with mass $M$ per mass interval $dM$, in the range $M_{\rm min}
< M < M_{\rm max}$, with $\int f(M)dM = f_{\rm tot}$.  We assume $M_{\rm max} \gg M_{\rm min}$.
Substituting these formulae, along with equation (\ref{lumdistance2})
into equation (\ref{rate1}), we obtain equation (\ref{rate2}).

\section{Discussion}

Both the distance reached by LISA and the estimated rate of inspirals
detected differ markedly from the estimates given by 
\citet{miller02}, namely $D_L \approx 200 (\mu/10 M_\odot)^{1/2}(M/100
M_\odot)^{1/2} \,{\rm Mpc}$, and equation (\ref{rate0}).  
The distance derived by Miller is larger and the mass dependence
is different than in equation (\ref{lumdistance2}) because he appears
to have  adopted an
equation [(B7)] from \citet{flanhughes} that actually
applies only to equal-mass systems.  
For the low masses relevant to IMBH in the year
leading up to the merger, the
high-frequency end of the LISA noise spectrum is dominant, where the
dependence on frequency (and hence on $M$) is steeply increasing.
But the underlying dependence is on {\em chirp} mass, so for a given total
mass, there is still a strong dependence on reduced mass parameter, which is not
reflected in equation [(B7)] of \citet{flanhughes}.
Also, the noise root spectral density
curve modeled by \citet{flanhughes} is lower by a
factor of about 3 -- 5 than that given by the SCG (figure 3).  Their
noise curve was based on an out-of-date description of the LISA
mission.
Distance estimates
using the Flanagan-Hughes curve will therefore automatically
be larger and rates
will be higher than those
using the SCG.  

We have assumed throughout that the inspiral orbits are quasi-circular, that
is, circular apart from the adiabatic inspiral due to radiation reaction.
In reality, IMBH binary orbits are likely to be highly eccentric
\citep{gultekin}.  This will increase the average gravitational-wave flux
for a given orbital period, and could add harmonics in frequency bands where
LISA's noise may be lower;  on the other hand it will increase the number of
parameters to be estimated in the matched filter, which raises the
threshold needed to achieve detection.
It is not clear whether the net effect of these complicated and possibly
offsetting effects will improve LISA's reach for IMBH binaries or not.  In
any event, they are beyond the scope of this paper.

\acknowledgments

We are grateful for useful discussions with Cole Miller and Scott Hughes, who
have confirmed the conclusions reached here.
We thank the Institut d'Astrophysique de Paris for its
hospitality during the academic year 2003-04.  This work was supported
in part by the National Aeronautics and Space Administration under
grant no. NAG 5-10186 and the National Science Foundation,
under grants no. PHY 00-96522 and PHY 03-53180.


\end{document}